\documentstyle[sprocl]{article}

\input{psfig}

\newcommand{\gsim}{\raisebox{-0.7ex}{$\stackrel{\textstyle >}{\sim}$ }}

\def\pislash{ {\pi\hskip-0.6em /} }
\def\pislashsmall{ {\pi\hskip-0.375em /} }

\def\nopi{ {\rm EFT}(\pislash) }
\def\si{{}^1\kern-.14em S_0}
\def\siii{{}^3\kern-.14em S_1}
\def\diii{{}^3\kern-.14em D_1}
\def\CA{{\cal A}}

\def\frac#1#2{{\textstyle{#1\over#2}}}
\def\darr#1{\raise1.5ex\hbox{$\leftrightarrow$}\mkern-16.5mu #1}
\def\){\right)}
\def\({\left(}
\def\]{\right]}
\def\[{\left[}
\def\MeV{{\rm\ MeV}}
\def\CA{{\cal A}}

\def\Czerominus{ {^\pislashsmall \hskip -0.2em C_{0,-1}^{(\siii)} }}

\def\CSDzero{ {^\pislashsmall \hskip -0.2em C_0^{(sd)} }}

\def\CSDtwotwo{ {^\pislashsmall \hskip -0.2em C_2^{(sd)} }}
\def\CSDtwotwotwo{ {^\pislashsmall \hskip -0.2em C_{2,-2}^{(sd)} }}
\def\CSDzeroone{ {^\pislashsmall \hskip -0.2em C_{0,-1}^{(sd)} }}
\def\CSDzerotwo{ {^\pislashsmall \hskip -0.2em C_{0,0}^{(sd)} }}
\def\Ltwo{ {^\pislashsmall \hskip -0.2em L_2 }}
\def\Lone{ {^\pislashsmall \hskip -0.2em L_1 }}
\def\CQuad{ {^\pislashsmall \hskip -0.2em C_{\cal Q} }}

\def\Journal#1#2#3#4{{#1} {\bf #2}, #3 (#4)}

\def\NPB{{\em Nucl. Phys.} B}
\def\NPA{{\em Nucl. Phys.} A}
\def\NP{{\em Nucl. Phys.} }
\def\PLB{{\em Phys. Lett.}  B}
\def\PRL{\em Phys. Rev. Lett.}

\def\PRC{{\em Phys. Rev.} C}
\def\PRA{{\em Phys. Rev.} A}
\def\PR{{\em Phys. Rev.} }

\def\FBS{{\em Few Body Systems Suppl.}}

\def\be{\begin{equation}}
\def\ee{\end{equation}}
\def\bea{\begin{eqnarray}}
\def\eea{\end{eqnarray}}

\begin{document}

\title{The Two-Nucleon Sector with Effective Field 
Theory\footnote{NT@UW-99-23}    }

\author{Martin J. Savage}

\address{Department of Physics, University of Washington,  
Seattle, \\ WA 98915, USA 
\\E-mail: savage@phys.washington.edu}

\maketitle\abstracts{
I present the results obtained for several observables in the
two-nucleon sector using effective field theory with KSW power-counting
and  dimensional regularization.
In addition to the phase shifts for nucleon-nucleon scattering,
several deuteron observables are discussed,
including electromagnetic form factors,
polarizabilities,
$\gamma d\rightarrow \gamma d$ Compton scattering,
and $np\rightarrow d\gamma$.
A detailed comparison between the effective field theory with pions,
the theory without pions, and effective range theory is made.
  }

\section{Introduction}

During the last year there has been a tremendous effort to understand
the role of effective field theory (EFT) in the two- and three-nucleon
sectors\cite{Weinberg1}$^-$\cite{CRSnopi}.
Since the previous
{\it Nuclear Physics with Effective Field Theory} workshop
held twelve months ago at Caltech
I have been primarily focused on exploring the implications of KSW power
counting\cite{KSW} in the two-nucleon sector.
In addition to the nucleon-nucleon scattering phase shifts in the
spin-singlet $\si$ and spin-triplet $\siii$ channels,
various properties of the deuteron and processes involving the
deuteron have been considered.

If one is interested in the static properties of the deuteron or very low-energy
($|{\bf p}| \ll m_\pi$)
nucleon-nucleon scattering the pion can be ``integrated'' out of the
theory, leaving a very simple EFT involving only nucleons and
external currents (we will denote this theory as $\nopi$).
The simplicity of the theory allows for many observables to be computed up to
next-to-next-to-leading order (NNLO) with ease\cite{CRSnopi}, and enables a
direct comparison with effective range theory\cite{ERtheory,Noyes} to be made.
The expansion parameter in $\nopi$  is $Q\sim|{\bf p}|/m_\pi$, which for
static properties of the deuteron corresponds to $Q\sim {1\over 3}$.
A direct comparison between $\nopi$  and effective range theory
shows that effective range theory reproduces EFT up
to the order at which multi-nucleon-external-current operators enter.
At that order and beyond, effective range theory fails to reproduce effective
field theory, and thus is an incomplete description of the strong interactions.

For processes involving higher momenta, such as $|{\bf p}| \gsim
m_\pi$, the pions must be included as  a dynamical field.
In KSW power-counting, the exchange of potential pions is a subleading
interaction compared to the momentum-independent, quark mass independent
four-nucleon operator, and is treated in perturbation theory.
Some have questioned the convergence of perturbative pions\cite{CohHan}, and we
will see that for the deuteron observables we have considered the subleading
contribution from potential pion exchange is smaller than the contribution from
higher derivative four-nucleon operators.
While this does not directly answer
the questions raised by Cohen and Hansen \cite{CohHan}, it is an important
observation.
For the static properties of the deuteron, one recovers the results of the $\nopi$
with corrections suppressed by terms higher order in $\nopi$, as expected.


\section{EFFECTIVE FIELD THEORY WITH PIONS}

\subsection{THE NUCLEON-NUCLEON INTERACTION}

The two-nucleon sector contains length scales that are much larger
than one would
naively expect from QCD.
Scattering lengths in the s-wave channels are $a^{(\si)} = -23.7\ {\rm fm}$
in the $\si$ channel and $a^{(\siii)} = 5.4\ {\rm fm}$
in the $\siii$ channel, much greater than both
$1/\Lambda_\chi\sim 0.2\ {\rm  fm}$
and $1/f_\pi\sim 1.5\ {\rm  fm}$, typical hadronic scales.
\begin{figure}[h]
\qquad\qquad\qquad\qquad\psfig{figure=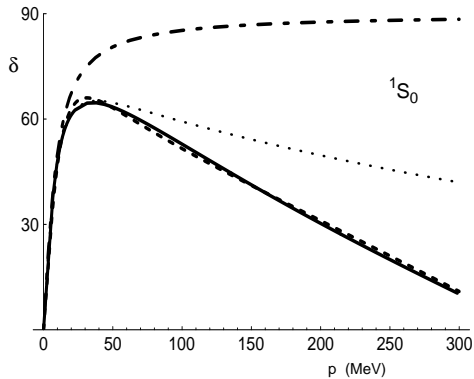,height=2.0in}
\caption{
The phase shift $\delta^{(\si)}$ verses the momentum of each
nucleon in the
center-of-mass.
The solid line denotes the Nijmegen partial wave analysis of the experimental
data.
\label{singphase}}
\end{figure}
The solid curve in fig.~(\ref{singphase}) shows the phase shift 
in the $\si$ channel, $\delta^{(\si)}$,
from the  Nijmegen partial wave analysis\cite{nijmegen} of the experimental data.
$\delta^{(\si)}$ rises very fast at low momentum and then turns over at about
$|{\bf p}|\sim 40\ {\rm MeV}$.

The lagrange density that describes the interaction between two
S-wave nucleons has the form
\begin{eqnarray}
  {\cal L} & = & -C_0^{(\si)} (N^T P^a N)^\dagger(N^T P^a N)
   - D_2^{(\si)} \omega\  Tr[ m_q ] (N^T P^a N)^\dagger(N^T P^a N)
   \nonumber\\
  & &  + {C_2^{(\si)}\over 8} \left[(N^T P^a N)^\dagger \left(N^T P^a
( \overrightarrow {\bf \nabla} - \overleftarrow {\bf \nabla} )^2 N\right) +
 h.c.\right]
\nonumber\\
 & & - g_A N^\dagger {\bf \sigma}\cdot {\bf A} N
 \ +\ (\siii)\ +\ ...
\label{eq:lagint}
\end{eqnarray}
where the ellipses denote terms with more insertions of
the small expansion parameters
$Q\sim m_q$ and $Q\sim \nabla$.
The reason for writing such an expansion is that one hopes or expects that QCD
will give rise to S-matrix elements for  multiple nucleon processes
that have a systematic expansion in the light quark masses and in
external momenta.
Only interactions in the  $\si$ channel are explicitly shown in
Eq.~(\ref{eq:lagint}),
while interactions in the $\siii$ channel are denoted by ``$(\siii)$''.
In addition, terms from $D_2$ involving pion field
operators (required by chiral symmetry), and
terms involving the photon field are not explicitly shown in Eq.~(\ref{eq:lagint}).
${\bf A}$ is the axial vector meson field.
The $P^a$ is a spin-isospin projector, $P^a = {1\over\sqrt{8}}\sigma_2
\tau_2\tau^a$, projects onto the $\si$ channel.

\begin{figure}[h]
\qquad\qquad\psfig{figure=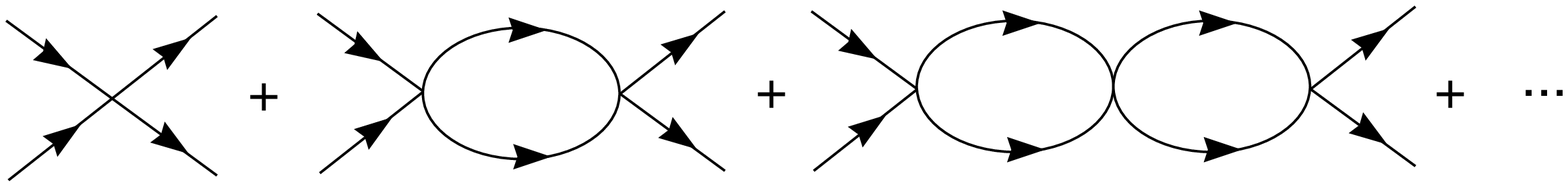,height=1.0in}
\caption{The leading contribution to nucleon-nucleon scattering
  arising from local operators.
\label{KSW_fig2}}
\end{figure}
\begin{figure}[h]
\qquad\qquad\psfig{figure=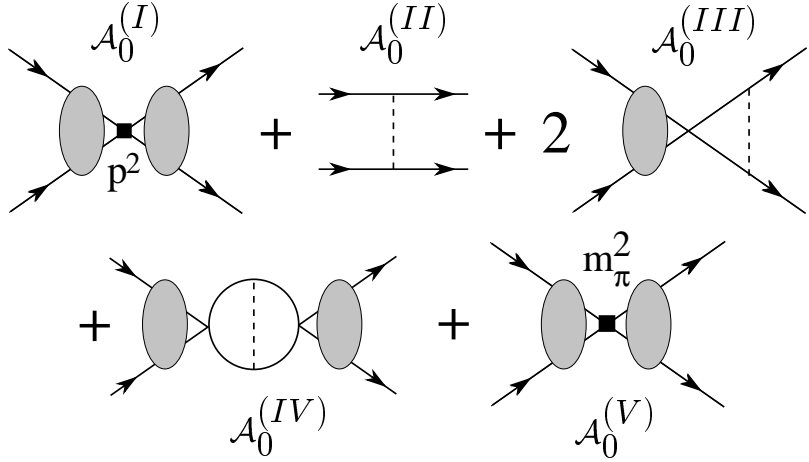,height=2.5in}
\caption{Graphs contributing to the subleading amplitude $\CA_{0}$.
The shaded ovals correspond to the bubble chain in Fig.~(\ref{KSW_fig2})
and free propagation. 
\label{KSW_fig5}}
\end{figure}

NN scattering in the $\si$ channel receives contributions at
leading order from the graphs shown in
fig.~(\ref{KSW_fig2}), i.e the bubble-chain of contact operators.
Choosing the coefficient $C_0^{(\si)}$ to reproduce the scattering length gives
the dot-dashed curve in fig~(\ref{singphase}).
At subleading order, there are contributions from the $C_2, D_2$ operators, in
addition to the exchange of a potential pion, as shown in
fig~(\ref{KSW_fig5}).
The best fit to the phase shift for momenta less than $\sim 250~{\rm MeV}$
is shown by the dashed curve in fig~(\ref{singphase}), while the fit that
reproduces both the scattering length and effective range is shown as the
dotted curve in fig~(\ref{singphase}).
\begin{figure}[h]
\qquad\psfig{figure=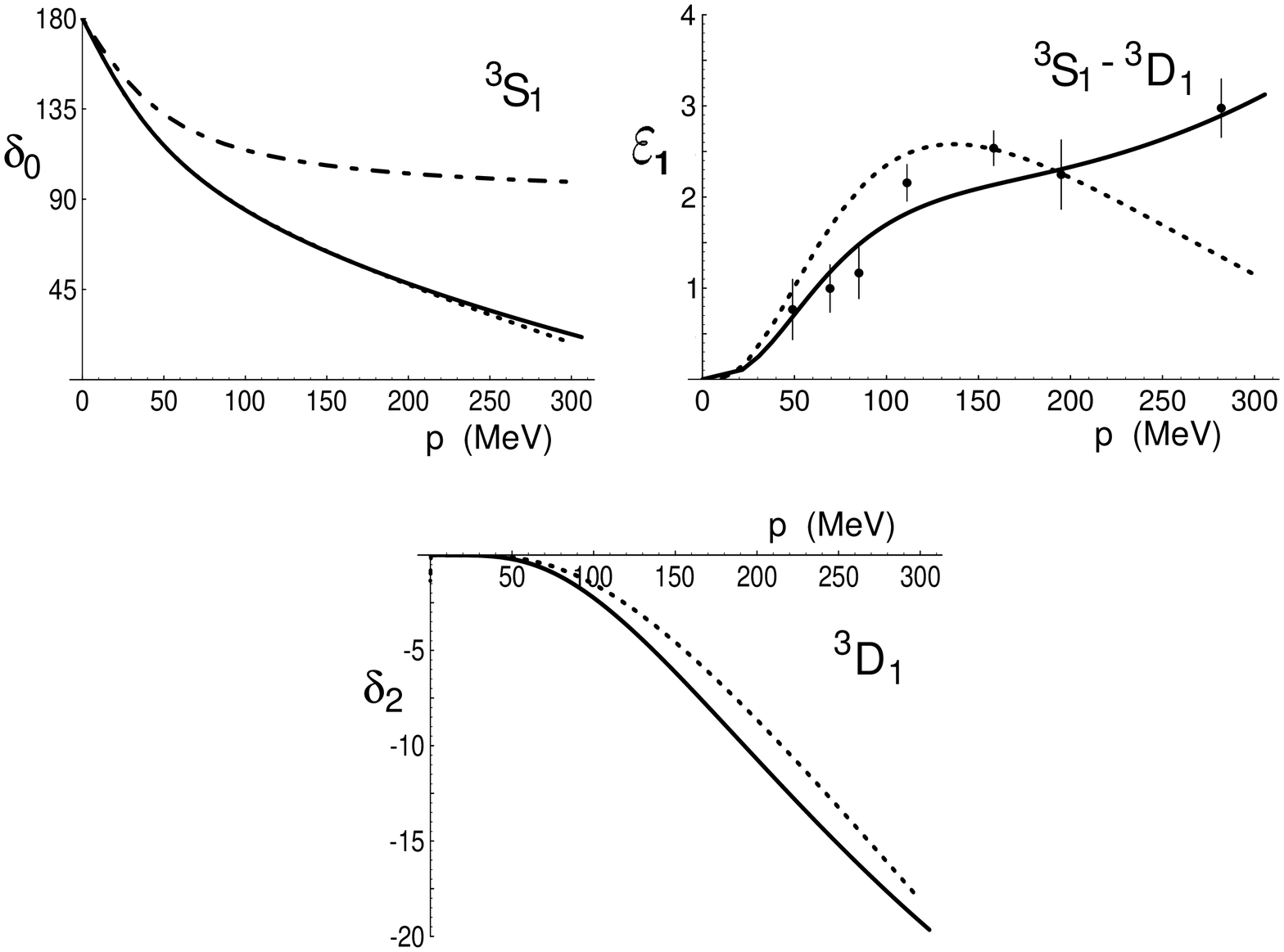,height=3.5in}
\caption{The phase shifts $\delta_0$, $\delta_2$ and  mixing parameter
$\overline{\varepsilon}_1$ for the $\siii-\diii$ channel. 
The solid line denotes the results of the  Nijmegen partial wave  analysis.
The dot-dashed curve is the fit at order $Q^{-1}$ for $\delta_0$, while
$\delta_2 = \overline{\varepsilon}_1 = 0$ at this order.
The dashed curves are the results of the order $Q^{0}$  fit of $\delta_0$ 
to the partial wave analysis over the momentum range $p\le 200\ \MeV$.
\label{KSW_fig7}}
\end{figure}
Analysis of scattering in the $\siii-\diii$ channel is a
straightforward extension of the analysis in the 
$\si$ channel.
The important difference is that the nucleons in the initial 
and final states with total angular momentum $J=1$ can be in an 
orbital angular momentum state of either $L=0$ or $L=2$.
The S-matrix describing scattering in the coupled channel $J=1$
system is written as
\begin{eqnarray}
  S & = & \left(
    \matrix{ e^{i2\delta_0}\cos 2\overline{\varepsilon}_1
      & i e^{i(\delta_0+\delta_2)}\sin2\overline{\varepsilon}_1
      \cr
       i e^{i(\delta_0+\delta_2)} \sin2\overline{\varepsilon}_1
      &
       e^{i2\delta_2}\cos 2\overline{\varepsilon}_1
}\right)
\ \ \  ,
\label{eq:Smat}
\end{eqnarray}
where we use the ``barred'' parameterization of \cite{stapp}, also used in
\cite{Nij}.
The power counting for amplitudes that take the nucleons from a 
$\siii$-state to a $\siii$-state is identical to the analysis
in the $\si$-channel.
Operators  between two $\diii$ states are not directly renormalized by the 
leading operators, which project out only $\siii$ states.
However, they are renormalized by operators that mix the $\siii$ and $\diii$ 
states, which in turn are renormalized by the leading interactions.
Further, they involve a total of four spatial derivatives, two on the incoming 
nucleons, and  two on the out-going nucleons.   
Therefore, such operators contribute
at order $Q^3$, and can be neglected in the present computation.
Consequently, amplitudes for scattering from an $\diii$ state into an $\diii$ state
are dominated by single potential pion exchange at order $Q^0$.
Operators connecting $\diii$ and $\siii$ states
are renormalized by the leading  operators, but only on the 
$L=0$ ``side'' of the operator.
Therefore the coefficient of this operator, $C_2^{(sd)}\sim Q^{-1}$,
indicates that this interaction
contributes at order $Q^1$ and can be neglected at order $Q^0$.
Thus,  mixing between $\diii$ and $\siii$ states is dominated by
single potential pion exchange dressed by a bubble chain of
$C_0^{(\siii)}$ operators.
A parameter free prediction for this mixing 
exists at order $Q^0$.
The dashed curves in Fig.~(\ref{KSW_fig7})  show the  phase shifts
$\delta_0$, $\delta_2$
and mixing parameter $\overline{\varepsilon}_1 $
compared to the  Nijmegen partial wave  analysis\cite{nijmegen}
for this set of coefficients.
There are no free parameters at this order in either
$\overline{\varepsilon}_1 $ or $\delta_2 $.

\subsection{THE DEUTERON}

\subsubsection{Electromagnetic Form Factors and Moments}

\begin{figure}[h]
\psfig{figure=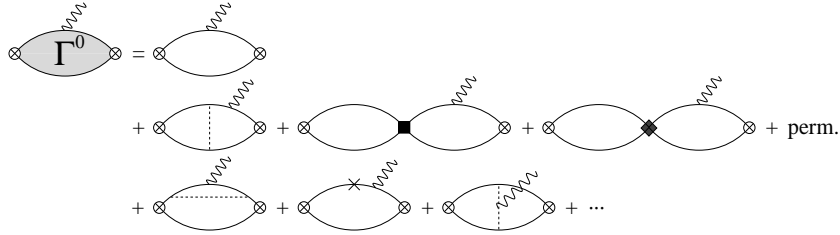,height=1.3in}
\caption{The diagrams contributing to the electric form factors of the deuteron.
\label{fig:gam0}}
\end{figure}
Once the Lagrange density in the nucleon sector has been
established the standard tools of field theory
can be used to determine the properties of
the deuteron\cite{KSW2}.
To compute the electromagnetic form factors of the deuteron one first computes
the three point correlation function between a source that creates a nucleon
pair in a $\siii$ state, a source that destroys a nucleon
pair in a $\siii$ state and a source that creates a photon.
After LSZ reduction and wavefunction renormalization one obtains the
electromagnetic form factors.
Leading order (LO),
next-to-leading (NLO) and
next-to-next-to-leading order (NNLO)
graphs contributing to the electric
form factors of the deuteron are shown in Fig.~(\ref{fig:gam0}).
\begin{figure}[h]
\qquad\qquad\psfig{figure=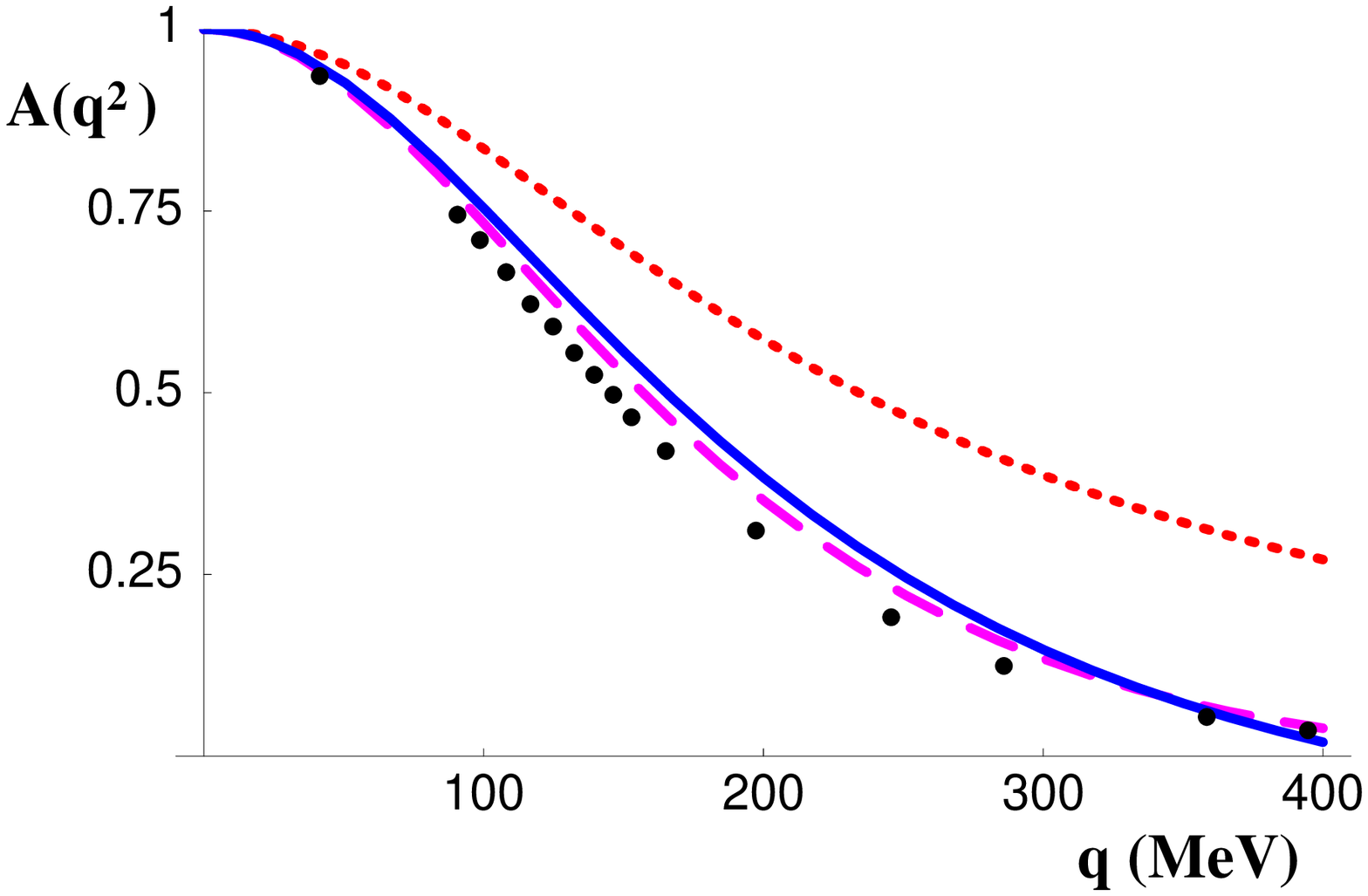,height=1.8in}

\qquad\qquad\psfig{figure=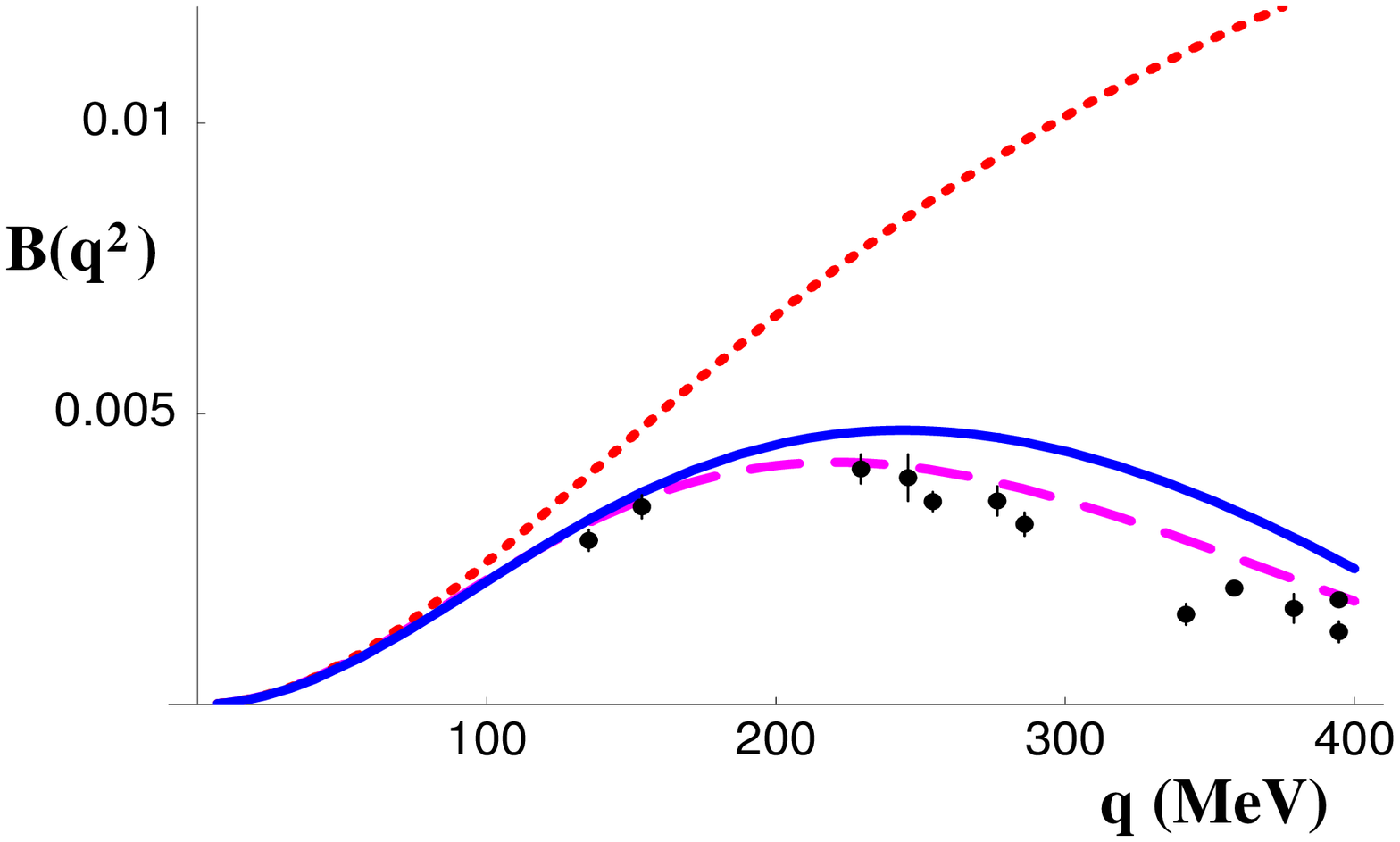,height=1.8in}
\caption{The form factors $A(q^2)$ and $B(q^2)$ measured in elastic
  electron-deuteron scattering.
  The dotted curve is the LO prediction, while the solid
  curve NLO prediction.
  The dashed curve is the prediction of effective range theory.
  There is one counterterm contributing to  $B(q^2)$ at this order
  which is fixed by the deuteron magnetic moment.
\label{fig:ABplot}}
\end{figure}
The resulting form factors $A(q^2)$ and $B(q^2)$ that appear in the differential
cross section for electron deuteron scattering are shown in
Fig.~(\ref{fig:ABplot}).
$A(q^2)$ is dominated by the charge form factor and $B(q^2)$ depends only upon
the magnetic form factor.
One sees that the form factors computed at subleading order agree well with the
data.
The charge radius of the deuteron is found to be
\begin{eqnarray}
  \sqrt{<r^2>}_{\rm th} & = & {1\over 2\sqrt{2}\gamma}
  \ +\  C_2(\mu) {M_N (\mu-\gamma)^2\over 8\sqrt{2} \pi}
\ +\
{g_A^2 M_N m_\pi^2 (3 m_\pi + 10\gamma)\over 48\sqrt{2}\pi f_\pi^2
    (m_\pi + 2 \gamma)^3}\ +\ ...
  \nonumber\\
  & = & 1.53\ \ +\ \ 0.36\ \ +\ \ ...\ \ =\ \ 1.89\  {\rm fm} + ...
  \nonumber\\
   \sqrt{<r^2>}_{\rm expt} & = & 2.11\  {\rm fm}
   \ \ \ ,
\label{eq:rtwo}
\end{eqnarray}
where $\gamma=\sqrt{M_N B}$ is the binding momentum of the deuteron.
In computing the numerical value we have used the numerical value of $C_2$
extracted from a fit to data over the range of momenta $|{\bf p}| < 250\ {\rm
  MeV}$.
Given the expansion parameter for the theory is $\sim {1\over 3}$, we expect
$ \sqrt{<r^2>}_{\rm th}$ given in Eq.~(\ref{eq:rtwo})
is within $\sim 10\%$ of the actual value,
which it is.

In effective range theory the electromagnetic form factors are assumed to
be dominated by the asymptotic S-wave deuteron wave function,
\begin{eqnarray}
  \psi^{({\rm ER})}({\bf r}) & = & \sqrt{{\gamma \over 2 \pi
 (1- \gamma r_0) }} {e^{-\gamma r} \over r}
  \ \ \ .
\end{eqnarray}
Assuming the small r part of the deuteron wave function is only important
for establishing the normalization condition, $F_C(0)=1$, the prediction of
effective range theory for the form factor $F_C(q^2)$ follows from the
Fourier transform of $ |\psi^{(ER)}({\bf r})|^2$,
\begin{eqnarray}
F^{(ER)}_C(q^2)=\left({1 \over 1- \gamma \rho_d} \right)
\left(\left(
{4 \gamma \over \sqrt{-q^2}}\right){\rm tan}^{-1}\left({\sqrt{-q^2} \over 4
  \gamma}\right) -\gamma\rho_d\right)
\ \ \ .
\label{eq:ERcff}
\end{eqnarray}
This yields a deuteron charge radius,
\begin{eqnarray}
  \sqrt{ \langle r_d^2\rangle^{\rm ER}}  & = &
  {1\over 2\sqrt{2}\gamma} {1\over \sqrt{1 - \gamma   \rho_d}}
   \nonumber\\
    & = & {1\over 2\sqrt{2}\gamma}
    \left[ 1 + {1\over 2} \gamma \rho_d\ + {3\over 8} \gamma^2 \rho_d^2\ +\
      ...\right]
    \ \ \ ,
\label{eq:ERcr}
\end{eqnarray}
which gives a numerical value of
$\sqrt{ \langle r_d^2\rangle^{\rm ER}}=1.98~{\rm fm}$,
very close to the
quoted value of the matter radius of the deuteron\cite{buch,FMS} of
$r_m=1.967\pm 0.002~{\rm  fm}$.
Conventionally, the charge radius is obtained by combining
the nucleon charge radius in quadrature with the matter radius, which agrees
very well with the experimental value.

Expanding the EFT expression for
$\sqrt{<r^2>}_{\rm th}$ in Eq.~(\ref{eq:rtwo})
in powers of $\gamma/m_\pi$, we have 
\begin{eqnarray}
\gamma \ll m_\pi &\rightarrow & 
 {1\over 2 \sqrt{2}\gamma}\left[ 1 + {1\over 2}\gamma r_0\ +\
     { g_A^2 M_N\gamma\over 4\pi f_\pi^2}{\gamma^2\over
       m_\pi^2}\ +\ ...\right]
\ \ \ ,
\end{eqnarray}
where $r_0$ is the effective range.
It is clear that the NLO calculation is the same order as a single insertion of
$r_0$, but there is also a contribution arising from pion exchange that
is beyond effective range theory.

The quadrupole moment vanishes at leading order in the expansion but receives
a contribution at subleading order from the exchange of one potential pion, giving
\begin{eqnarray}
  \mu_{{\cal Q}, th} & = & {g_A^2 M_N( 6 \gamma^2 + 9m_\pi\gamma + 4
    m_\pi^2)\over 30\pi f_\pi^2 (m_\pi + 2 \gamma)^3}
  \ + \ ....
  \nonumber\\
  & = & 0.40\ {\rm fm^2}\ + ... 
    \nonumber\\
  \mu_{{\cal Q}, expt} & = &  0.2859\ {\rm fm^2} 
\ \ \ ,
\end{eqnarray}
which is approximately $30\%$ larger than the experimental value.
Clearly, a NNLO calculation\cite{Binger}
is needed to ensure that the EFT
value is indeed converging to the experimental value.

In contrast, there is a local counterterm
contributing to the magnetic moment at NLO,
\begin{eqnarray}
\mu_M & = & \mu_p\ +\ \mu_n\ +\ L_2 (\mu) {\gamma\over 2\pi} (\mu-\gamma)^2\ +\ ...
\nonumber\\
& = & 0.88\ -\  0.02\ \ \  (fit) 
\ \ \ \ ,
\end{eqnarray}
which determines the counterterm $L_2$ at the scale $\mu$.
Once this counterterm has been determined from the deuteron magnetic moment, the
entire form factor $B(q^2)$ is determined to NLO.

\subsubsection{Polarizabilities}

\begin{figure}[h]
\qquad\qquad\qquad\psfig{figure=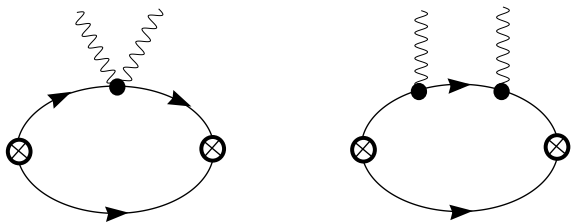,height=1.2in}
\caption{The leading diagrams contributing to the electric polarizability of
  the deuteron.
  Crossed graphs are not shown.}
\label{fig:pol}
\end{figure}
The scalar $\alpha_{E0}$ and tensor $\alpha_{E2}$ electric
polarizabilities are found to be\cite{CGSSpol}
\begin{eqnarray}
  \alpha_{E0} & = & {\alpha M_N\over 32 \gamma^4}\ +\ 
{\alpha M_N^2\over 64\pi\gamma^3} 
  C_2(\mu) (\mu-\gamma)^2
  \nonumber\\
& + & {\alpha g_A^2 M_N^2\over 384\pi f^2}\ 
  { m_\pi^2 ( 3 m_\pi^2 + 16 m_\pi\gamma + 24\gamma^2)\over
      \gamma^3 (m_\pi+2 \gamma)^4}
    \ +\ ...
  \nonumber\\
\alpha_{E2} & = & - {\alpha g_A^2 M_N^2\over 80\pi f^2} \ 
  { 2m_\pi^3+11 m_\pi^2\gamma + 16 m_\pi\gamma^2 + 8\gamma^3\over \gamma^2
    (m_\pi+2\gamma)^4}
  \nonumber\\
 \alpha_{E0} & = & 0.595\ {\rm fm}^3\ +\ ...\ \ ,\ \
 \alpha_{E2} =  -0.062\ {\rm fm}^3\ +\ ...
\ \ \ .
\label{eq:polex}
\end{eqnarray}
Numerically, the value of  $ \alpha_{E0}$ in Eq.~(\ref{eq:polex}) is expected
to be within $\sim 10\%$ of the actual value.
Potential models, which find a value of\cite{FriP}
$ \alpha_{E0}=0.632\pm 0.003\ {\rm fm}^3$,
are expected to reproduce this observable to high
precision as the short-distance contributions are expected to be small.

It is informative to perform a
momentum expansion of  $ \alpha_{E0}$ to make contact with effective range
theory,
\begin{eqnarray}
  \alpha_{E0} 
 & \rightarrow &
{\alpha M_N\over 32\gamma^4}
  \left[
    1
    \ +\ C_2(\mu)\ {M_N\gamma (\mu-\gamma)^2\over 2\pi }
\right.\nonumber\\
& & \left.
  \qquad\qquad
  \ +\ { g_A^2 M_N\gamma \over 4 \pi f^2} 
    \left( 1 - {8 \over 3}{\gamma\over m_\pi} + {16\over 3}
      {\gamma^2\over m_\pi^2}\ +\ ...\right) 
    \right]
\nonumber\\
 & = &
{\alpha M_N\over 32\gamma^4}
  \left[ 1 \ +\  \gamma r_0
  \ +\ { g_A^2 M_N\gamma \over 4 \pi f^2}{10\over 3}
      {\gamma^2\over m_\pi^2}\ +\ ...
      \right] 
\ \ \ .
\label{eq:polER}
\end{eqnarray}
The momentum expansion of Eq.~(\ref{eq:polex}) agrees with 
effective range theory up to order $\gamma^2$ at linear order in $r_0$.
The deviations from effective range theory
arise from the electromagnetic interactions of the pions, which
are beyond effective range theory.


\subsubsection{$\gamma$-Deuteron Compton Scattering}

\begin{figure}[h]
\qquad\qquad\qquad\qquad\psfig{figure=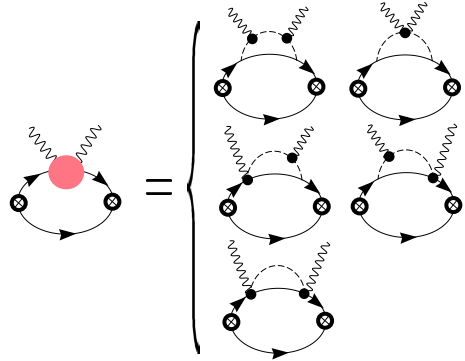,height=2.0in}
\noindent
\caption{\it Pion graphs that contribute to the nucleon
polarizabilities and to $\gamma$-deuteron
Compton scattering at NLO.
  The crossed circles denote operators that create or
  annihilate two nucleons with the quantum numbers
  of the deuteron.
  The dark solid circles correspond to the photon
  coupling via the nucleon or pion kinetic energy
  operator or via the gauged axial pion-nucleon interaction.
The solid lines are nucleons and the dashed lines
are pions.
The photon crossed graphs are not shown.
}
\label{fig:Polariz}
\vskip .2in
\end{figure}
A process that may allow for a determination of the neutron polarizabilities
(given the proton polarizabilities)
is $\gamma$-deuteron Compton scattering.
In power counting the graphs that contribute to the deuteron polarizability,
it was assumed\cite{CGSSpol}
that $E_\gamma\sim |{\bf p}_\gamma|\sim Q^2$.
However, for Compton scattering at photon energies of $\sim 100\ {\rm MeV}$,
the appropriate scaling is $E_\gamma\sim |{\bf p}_\gamma|\sim Q$.
With this power counting (called Regime II counting\cite{CGSSpol}), the
nucleon isoscalar polarizabilities contribute at NLO through the graphs shown
in fig~(\ref{fig:Polariz}).

\begin{figure}[h]
\qquad\qquad\qquad\psfig{figure=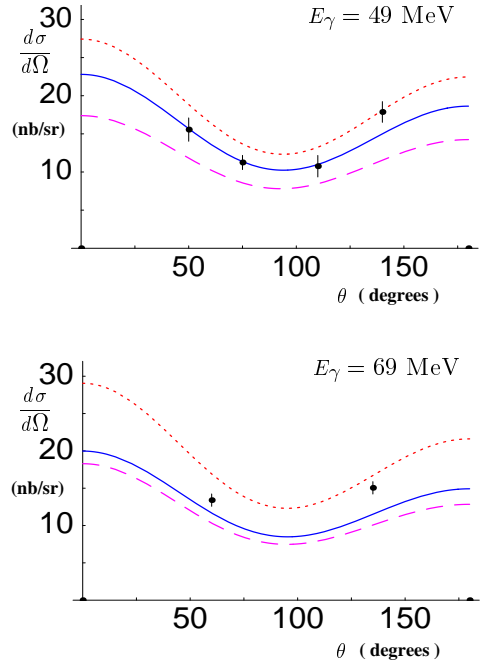,height=3.5in}
\noindent
\caption{\it The differential cross section for
  $\gamma$-deuteron Compton scattering at incident
  photon
  energies of
  $E_{\gamma}=49\ {\rm MeV}$  and  $69\  {\rm MeV}$.
  The dashed curves correspond to the  LO result.
  The dotted curves correspond to the NLO result without the graphs that
  contribute to the polarizability of the nucleon.
  The solid curves correspond to the complete NLO result
  with no free parameters, as described in the text.
  Systematic and statistical errors associated with each data point 
  have been added in quadrature.
  }
\label{fig:diffcr}
\vskip .2in
\end{figure}

The cross section for $\gamma$-deuteron Compton
scattering\cite{CGSSpol} up to NLO is shown in fig~(\ref{fig:diffcr})
for incident photon energies of $E_\gamma~=~49~{\rm MeV}$ and $69~{\rm MeV}$.
There are no published data for this cross section,
but there is unpublished data
in the PhD Thesis of M.~Lucas\cite{Lucas}, consisting of four data points at
$E_\gamma=49\ {\rm MeV}$ and two data points at $E_\gamma=69\ {\rm MeV}$.
In fig~(\ref{fig:diffcr}) the dashed curve is the  LO calculation
which is consistently below the data.
The dotted curve is the NLO calculation
with the nucleon polarizabilities resulting from the graphs in
fig~(\ref{fig:Polariz}) set equal to zero.
This overestimates the cross
section, with the exception of the back angle data point at
$E_\gamma=49\ {\rm  MeV}$.
The solid curve is the complete NLO calculation which agrees reasonably well
with the data.

There are two published potential model calculations of this cross
section\cite{LLa,WWA}.
Their results disagree with each other at the $\sim 10-15 \%$ level, and do not
reproduce the data at both energies.
A more sophisticated potential model calculation
has recently been completed by Karakowski and Miller\cite{KarMil}.
Further, a calculation of this process is being performed using Weinberg's
power counting by Beane, Philips and van Kolck\cite{BPv}.
In the near future more data is expected to become available
which will allow for a clearer picture of our understanding of this process.


\subsubsection{$np\rightarrow d\gamma$}

One of the classic nuclear physics processes is the radiative capture
$np\rightarrow d\gamma$.
It is a clear demonstration of the existence of 
 ``meson exchange currents'' in nuclei.
The cross section for this process has been measured long ago, and for neutrons
incident in the lab with speed $|{\bf v}|=2200\ {\rm m/s}$, the cross section
is  $\sigma^{\rm expt} = 334.2\pm 0.5\ {\rm mb}$\cite{CWCa}.
The theoretical cross section for this process has been
calculated with effective
range theory\cite{ERtheory,Noyes}, 
and also more recently using a potential model
motivated by Weinberg's power counting\cite{Parka}.

The amplitude for
the radiative capture of extremely low momentum neutrons
$np\rightarrow d\gamma$
has contributions from both the $\si$ and $\siii$ $NN$
 channels. It can be written as
\begin{eqnarray}
\label{eq:matrix}
i{\cal A} & = &
e\ X\ N^T\tau_2\ \sigma_2 \  \left[ {\bf \sigma}\cdot {\bf k}\ 
\ {\bf\epsilon} (d)^* \cdot {\bf \epsilon} (\gamma)^*
  \ -\ {\bf \sigma} \cdot  {\bf \epsilon} (\gamma)^*\ 
  \ {\bf k}\cdot {\bf \epsilon} (d)^* 
  \right] N 
\\ \nonumber
& + &
i e\ Y\  \epsilon^{ijk}\ \epsilon (d)^{i*}\   
k^j\  {\bf\epsilon} (\gamma)^{k*}
\ (N^T\tau_2 \tau_3 \sigma_2 N)
\ \ \ \ ,
\end{eqnarray}
where  $e=|e|$ is the magnitude of the  electron charge, 
$N$ is 
the doublet of nucleon spinors, ${\bf \epsilon}(\gamma)$ is 
the polarization vector for the photon, ${\bf \epsilon}(d)$ is the polarization
vector for the deuteron and ${\bf k}$ is the outgoing photon momentum.
The term with coefficient $X$ corresponds to capture from the $\siii$ channel
while the term with coefficient $Y$ corresponds to capture from the $\si$
channel.
For convenience, we define dimensionless variables $\tilde X$ and $\tilde Y$,
by
\begin{eqnarray}
  X & = & i {2\over M_N} \sqrt{\pi\over\gamma^3}\ \tilde X
  \ \ ,\ \ 
  Y =  i {2\over M_N} \sqrt{\pi\over\gamma^3}\ \tilde Y
  \ \ \ \ .
\end{eqnarray}
Both $\tilde X$ and $\tilde Y$ have the $Q$ expansions,
$\tilde X = \tilde X_0+ \tilde X_1+...$, and $\tilde Y=\tilde Y_0+ \tilde
Y_1+...$,
where a subscript $n$ denotes a contribution of order $Q^n$.

\begin{figure}[h]
\qquad\qquad\qquad\qquad\psfig{figure=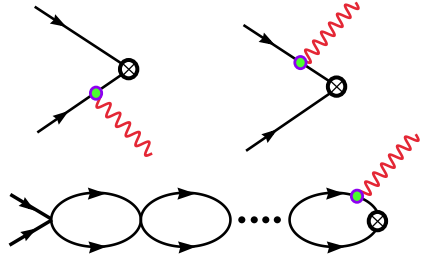,height=1.4in}
\noindent
\caption{\it Graphs contributing to the amplitude  
  for $np\rightarrow d\gamma$ at leading order in the
  effective field theory expansion. 
  The solid lines denote nucleons
  and the wavy lines denote photons.
The light solid circles correspond to the
nucleon magnetic
moment coupling to
the electromagnetic field.
  The crossed circle represents an insertion of
  the deuteron
  interpolating 
  field which is taken to have $\siii$ quantum numbers.
}
\label{fig:npstrong}
\vskip .2in
\end{figure}
At leading order in the EFT\cite{SSWst}
the graphs shown in fig~(\ref{fig:npstrong})
give a contribution to the cross section of
\begin{equation}
\label{eq:npleading}
\sigma^{LO}={8\pi\alpha\gamma^5\kappa_1^2 (a^{(\si)})^2 \over |{\bf v}| M_N^5}
\left(1-{1 \over \gamma a^{(\si)}} \right)^2
\ =\ 297.2\ {\rm mb}
\ \ \  .
\end{equation}
This agrees with the effective range theory calculation
of Bethe and Longmire~\cite{ERtheory} and Noyes~\cite{Noyes}
when terms in their expression involving the effective range are neglected.
Eq.~(\ref{eq:npleading}) is about $10\%$ less than the experimental value,
$\sigma^{\rm expt} = 334.2\pm 0.5\ {\rm mb}$\cite{CWCa}.

At NLO there are contributions from insertions of the $C_2, D_2$ operators and
from potential pion exchange with the isovector nucleon magnetic moment.
\begin{figure}[h]
\qquad\qquad\qquad\qquad\qquad\psfig{figure=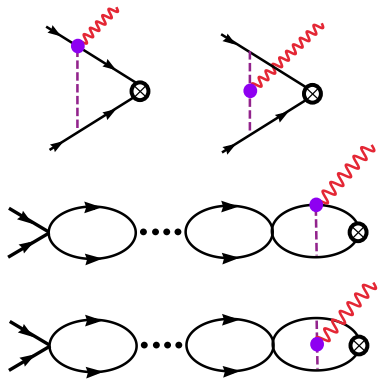,height=2.0in}
\noindent
\caption{\it Pion exchange current contributions
  to the amplitude for $np\rightarrow d\gamma$ arising at NLO. 
  The solid lines denote nucleons
  and the wavy lines denote photons.
  The dashed line denotes a pion.
The dark solid circles correspond to minimal
coupling of the photon.
The crossed circle represents an insertion of
the deuteron interpolating field. }
\label{fig:strongsubpiE}
\vskip .2in
\end{figure}
\begin{figure}[h]
\qquad\qquad\qquad\psfig{figure=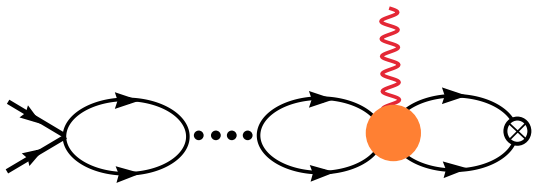,height=1.0in}
\noindent
\caption{\it Local counterterm contribution to the
  amplitude  
  for $np\rightarrow d\gamma$ at NLO.
  The solid lines denote nucleons
  and the wavy lines denote photons.
  The solid circle corresponds to an
  insertion of the $L_1$ operator.
  The crossed circle represents an insertion
  of the deuteron interpolating field.
  The graphs corresponding to final state
  interactions are not shown.}
\label{fig:strongsubL1}
\vskip .2in
\end{figure}
In addition, there are contributions from meson exchange where the photon is
minimally coupled either to the axial current or to the pion itself.
These diagrams, shown in fig~(\ref{fig:strongsubpiE}), are called
``meson exchange currents''.
Also, at this order there is a contribution from a four-nucleon-one-photon
counterterm, shown in fig~(\ref{fig:strongsubL1}).
The contributions from the meson exchange currents and the
counterterm\cite{SSWst} are
\begin{eqnarray}
\tilde Y^{(\rm \pi, E)} & = & {g_A^2 M_N \gamma ^2 \over 12\pi f^2} 
  \left( {m_\pi-\gamma\over (m_\pi+\gamma)^2}
    \right.\nonumber\\
 & & \left. \qquad\qquad
    \ +\
    { M_N \over 4\pi}{\cal A}_{-1}^{(\si)}(0)
    \left( { 3m_\pi-\gamma\over 2(m_\pi+\gamma)} +
      {\rm ln}\left({m_\pi+\gamma\over\mu}\right) - {1\over 6} + \delta
    \right)\right)
  \nonumber\\
  \tilde Y^{(\rm L_1)} & = &  L_1\ \gamma^2\ { {\cal
      A}_{-1}^{(\si)}(0)\over \bar C_0^{(\si)}
     \bar C_0^{(\siii)}}
  \ \ \ ,
\label{ys}
\end{eqnarray}
where $\delta$ is a subtraction constant.
The important point to note here is that the contribution from the meson
exchange currents is UV divergent,  as is clear from the presense of the
$\log(\mu)$ contribution in eq.~(\ref{ys}), and the subtraction constant
$\delta$.  This divergence is absorbed by the local counterterm $L_1 (\mu)$,
and  therefore it makes no sense to discuss the contribution from meson exchange
currents alone.

The two four-nucleon-one-photon counterterms that arise at NLO are renormalized
very differently.  In contrast to the counterterm
$L_1 (\mu)$, the counterterm $L_2 (\mu)$ is not driven by meson exchange
currents and satisfies a simple renormalization group equation,
\begin{eqnarray}
   \mu {d\over d\mu}
  \left[  { L_1 - {1\over 2}\kappa_1 \left( C_2^{(\si)} +
        C_2^{(\siii)}\right)\over
       \bar C_0^{(\si)} \bar C_0^{(\siii)}}\right] 
      & = & 
       {g_A^2 M^2\over 48\pi^2 f^2}
\nonumber\\
 \mu {d\over d\mu}
    \left[ { L_2 \over (\bar C_0^{(\siii)})^2} \right] 
      & = & 0 
\ \ \ .
\end{eqnarray}


\section{EFFECTIVE FIELD THEORY WITHOUT PIONS}

For processes involving external momenta much less than the pion mass,
it is much simpler to work with an EFT where the pions do not
appear.
In this pionless theory\cite{CRSnopi} ($\nopi$),
one can examine each of the observables we have just
discussed in the theory with dynamical pions.
In many of the cases the calculation can be pushed to higher order with little
effort, which makes the pionless theory particularly attractive for static
properties.
Without  pions the only expansion parameter is the external momentum,
$Q\sim {\bf p}$, where
we will neglect the electromagnetic interaction and isospin breaking\cite{Epel}.

\subsection{THE NUCLEON-NUCLEON INTERACTION}

\begin{figure}[h]
\qquad\qquad\qquad\qquad\qquad\psfig{figure=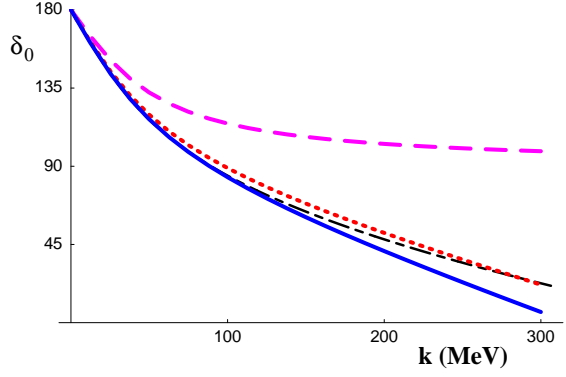,height=2.0in}
\noindent
\caption{\it The phase shift in the $\siii$ channel,
$\delta_0$, as a function of the 
center of mass momentum $|{\bf k}|$. 
The dashed curve corresponds to $\delta_0^{(0)}$, 
the dotted curve corresponds to $\delta_0^{(0)}+\delta_0^{(1)}$,
the solid curve corresponds to $\delta_0^{(0)}+\delta_0^{(1)}+\delta_0^{(2)}$,
and the dot-dashed curve is the Nijmegen partial wave
analysis\protect\cite{nijmegen}.
}
\label{fig:Sphase}
\vskip .2in
\end{figure}

In $\nopi$, the mixing parameter $\overline{\varepsilon}_1$ is supressed by
$Q^2$ compared with $\delta_0^{(0)}$, 
and therefore, up to N$^4$LO, we can isolate the
S-wave from the D-wave in the deuteron channel, leaving
\begin{eqnarray}
  S_{00} & = &  e^{i2\delta_0}\ =\ 1\ +\  {2 i\over \cot\delta_0 - i}
\ \ \ \ .
\label{eq:Smat0}
\end{eqnarray}
The phase shift $\delta_0$ has an expansion in powers of $Q$,
$\delta_0=\delta_0^{(0)}+\delta_0^{(1)}+\delta_0^{(2)}+~...$, where the
superscript denotes the order in the $Q$ expansion.
By forming
the logarithm of both sides of eq.~(\ref{eq:Smat0}) and
expanding in powers of $Q$, it is straightforward to obtain
\begin{eqnarray}
  \delta_0^{(0)} (|{\bf k}|) & = &
  \pi - \tan^{-1}\left({ |{\bf k}|\over \gamma}\right)
  \nonumber\\
  \delta_0^{(1)}(|{\bf k}|) & = & -{\rho_d\over 2}  |{\bf k}|
  \nonumber\\
  \delta_0^{(2)}(|{\bf k}|) & = &
  -\left[ {\rho_d^2\gamma\over 4}  + {\gamma^3\over 8 M_N^2
    (\gamma^2+ |{\bf k}|^2)}\right] |{\bf k}|
  \ \ \ ,
  \label{eq:phaseexp}
\end{eqnarray}
which are shown in fig.~(\ref{fig:Sphase}).
%
Up to NNLO, the shape parameter term,
$w_2$ does not contribute to the S-wave phase shift and
therefore, in addition to
being numerically small (about a factor of 5 smaller than $\rho_d$), $w_2$
enters only at higher orders.
Formally, the first deviations from linear
$|{\bf k}|$ dependence at large momenta arises from relativistic corrections
(the second term in $\delta_0^{(2)}$ in eq.~(\ref{eq:phaseexp}).
The $Q$ expansion is clearly demonstrated in eq.~(\ref{eq:phaseexp}),
by simply counting powers of $|{\bf k}|$ and $\gamma$,
which both scale like $Q$.
We expect this perturbative expansion of the phase shifts to converge
up to momenta of order $\sim m_\pi/2\sim 70~{\rm MeV}$,
at which point one encounters the t-channel cut from potential pion exchange.

Scattering between the S-wave and D-wave is induced by local operators first
arising at $Q^1$ in the power counting.
At order $Q^1$ and $Q^2$, the lagrange density describing such interaction is 
\begin{eqnarray}
  {\cal L}^{(sd)}_2 & = &
  {1\over 4} \CSDzero \left( N^T P^i N\right)^\dagger
  \left( N^T {\cal O}^{(sd) , xyj} N\right)
{\cal T}^{ijxy}\ +\ {\rm h.c.}
\nonumber\\
& - & {1\over 16}\CSDtwotwo
\left( N^T \left[ P_i \overrightarrow {\bf D}^2 +\overleftarrow {\bf D}^2 P_i
- 2 \overleftarrow {\bf D} P_i \overrightarrow {\bf D} \right]  N\right)^\dagger
  \left( N^T {\cal O}^{(sd) , xyj} N\right)
{\cal T}^{ijxy}\ +\ {\rm h.c.}
\ \ \ ,
\label{eq:sdlag}
\end{eqnarray}
where
\begin{eqnarray}
{\cal O}^{(sd) , xyj} & = &
      \overleftarrow {\bf D}^x  \overleftarrow {\bf D}^y P^j
+ P^j \overrightarrow {\bf D}^x  \overrightarrow {\bf D}^y
- \overleftarrow {\bf D}^x P^j\overrightarrow {\bf D}^y
-\overleftarrow {\bf D}^y P^j\overrightarrow {\bf D}^x
\nonumber\\
{\cal T}^{ijxy}
& = & 
\left( \delta^{ix}\delta^{jy} - {1\over n-1} \delta^{ij}\delta^{xy}\right)
\ \ \ ,
\label{eq:sdop}
\end{eqnarray}
and where $n$ is the number of spacetime dimensions.
The lagrange density in eq.~(\ref{eq:sdop}) appears somewhat complex.
However, when the electromagnetic field is ignored, the operator collapses to
\begin{eqnarray}
{\cal O}^{(sd) , xyj}&\rightarrow &
P^j  \left( \overleftarrow {\bf \nabla} - \overrightarrow {\bf \nabla}\right)^x
 \left( \overleftarrow {\bf \nabla} - \overrightarrow {\bf \nabla}\right)^y
\ \ \ ,
\end{eqnarray}
explicitly Galilean invariant.

The coefficients that appear in 
eq.~(\ref{eq:sdop})
themselves have an expansion in powers of $Q$,
e.g. $\CSDzero = \CSDzeroone + \CSDzerotwo+...$.
Performing a $Q$ expansion on the mixing parameter
$\overline{\varepsilon}_1 =
\overline{\varepsilon}^{(2)}_1
+\overline{\varepsilon}^{(3)}_1+...$
it is straightforward to demonstrate that
\begin{eqnarray}
\overline{\varepsilon}_1^{(2)} (|{\bf k}|) & = &
 {\sqrt{2}\over 3} \left({  \CSDzeroone \over \Czerominus  }\right)
{ |{\bf k}|^3\over\sqrt{\gamma^2+|{\bf k}|^2}}
\nonumber\\
& = &
-{M_N\over 4\pi} (\mu-\gamma)  {\sqrt{2}\over 3}\ 
\CSDzeroone  { |{\bf k}|^3\over\sqrt{\gamma^2+|{\bf k}|^2}}
\ \ \  .
\label{eq:eonelead}
\end{eqnarray}
Renormalization group invariance of this leading order contribution to
$\overline{\varepsilon}_1$ indicates that
$\CSDzeroone\propto (\mu-\gamma)^{-1}$.
Renormalizing at the scale $\mu=m_\pi$, and fitting to the Nijmegen phase shift
analysis\cite{nijmegen}, we find $\CSDzeroone=-4.57\ {\rm fm}^4$.

\begin{figure}[t]
\qquad\qquad\qquad\psfig{figure=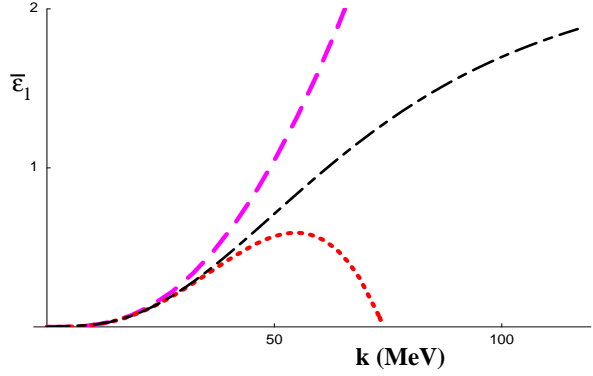,height=2.0in}
\noindent
\caption{\it The S-D mixing parameter $\overline{\varepsilon}_1$
as a function of the center of mass momentum $|{\bf k}|$. 
The dashed curve corresponds to $\overline{\varepsilon}_1^{(2)}$, 
the dotted curve corresponds to
$\overline{\varepsilon}_1^{(2)}+\overline{\varepsilon}_1^{(3)}$,
and the dot-dashed curve is the Nijmegen partial wave analysis.
}
\label{fig:Epphase}
\vskip .2in
\end{figure}

At the next order, $Q^3$, the contribution to
$\overline{\varepsilon}_1$ is
\begin{eqnarray}
 \overline{\varepsilon}_1^{(3)}(|{\bf k}|) & = &  -{\sqrt{2}\over 3} {M_N \over 4\pi} 
 \left[
\left( {\mu\gamma\rho_d \over 2}  \ \CSDzeroone
  \ +\  (\mu-\gamma)\ \CSDzerotwo \right)
{ |{\bf k}|^3\over\sqrt{\gamma^2+|{\bf k}|^2}}
\right. \nonumber\\
& & \left.   
\qquad   \ +\
\left( {\rho_d\over 2}\   \CSDzeroone + (\mu-\gamma)\  \CSDtwotwotwo\ \right)
{ |{\bf k}|^5\over\sqrt{\gamma^2+|{\bf k}|^2}}
   \right]
\ \ \ ,
\label{eq:eonesub}
\end{eqnarray}
where we have expanded $\CSDtwotwo=\CSDtwotwotwo+...$.
The first term in eq.~(\ref{eq:eonesub}) has the same momentum dependence as,
$\overline{\varepsilon}_1^{(2)}$ the
leading contribution in eq.~(\ref{eq:eonelead}).
In the same way that we have required the position of the deuteron pole is not
modified by higher orders in perturbation theory,
we may require that the coefficient contributing to $\overline{\varepsilon}_1^{(2)}$
is not modified by higher order contributions.  This constraint leads to
\begin{eqnarray}
 \CSDzerotwo & = &  -{\mu\gamma\rho_d \over 2 (\mu-\gamma)}  \ \CSDzeroone
\ \ \ .
\label{eq:csdtwocon}
\end{eqnarray}
Further, as the second term in eq.~(\ref{eq:eonesub})
is an observable, we obtain the RG equation
\begin{eqnarray}
\mu {d\over d\mu} 
\left( {\rho_d\over 2}\   \CSDzeroone + (\mu-\gamma)\  \CSDtwotwotwo\  \right)
& = & 0
\ \ \ ,
\label{eq:eoneRG}
\end{eqnarray}
The only free parameter
$\CSDtwotwotwo=+41.0~{\rm fm}^6$ (renormalized at $\mu=m_\pi$)
is fit to the
Nijmegen phase shift analysis\cite{nijmegen},
as shown in fig.~(\ref{fig:Epphase}).

%
%

\subsubsection{Electric Polarizability of the Deuteron}

With the $\nopi$ we have computed  $\alpha_{E0}$ up to NNLO, including
relativistic corrections.
It has a perturbative expansion in $Q$, and we write
 $\alpha_{E0} = \alpha_{E0}^{(-4)}+ \alpha_{E0}^{(-3)} +
 \alpha_{E0}^{(-2)}+...$.
Explicit calculation gives
\begin{eqnarray}
     \alpha_{E0}^{(-4)}+ \alpha_{E0}^{(-3)} + \alpha_{E0}^{(-2)}
     & = & 
{\alpha M_N\over 32\gamma^4}
    \left[ 1 \ \ +\ \  \gamma \rho_d\ \ +\ \  \gamma^2 \rho_d^2
      \ \ +\ \ {2\gamma^2\over 3 M_N^2}
      \right]
\nonumber\\
& = & \qquad 0.377\ +\ 0.153\ +\ 0.062\ +\ 0.0006\
\nonumber\\
& = & 0.592~{\rm fm}^3
\ \ \ .
\label{eq:eftpolnopi}
\end{eqnarray}
Numerically, the relativistic corrections are very small, two orders of
magnitude smaller than the NNLO corrections from the four-nucleon
interactions.
The value of $\alpha_{E0}$ in Eq.~(\ref{eq:eftpolnopi})
is within $\sim 5\%$ of that computed with potential models
and with effective range theory.
Despite being numerically small, relativistic corrections can be calculated
easily with the EFT.
The S-wave-D-wave mixing operators
(corresponding to the D-wave component of the deuteron in potential model
language)
make contributions to   $\alpha_{E}^{(-2)}$.
They do not contribute to the scalar polarizability,
$\alpha_{E0}^{(-2)}$, but do  contribute to the 
tensor polarizability, $\alpha_{E2}^{(-2)}$. 
Such operators will contribute to $\alpha_{E0}$ 
at higher orders in the expansion.


\subsubsection{Electromagnetic Form Factors of the Deuteron}

The charge radius of the deuteron at NNLO one finds in $\nopi$  is
\begin{eqnarray}
 \langle r_d^2\rangle^{\rm EFT} & = &\langle r_{N,0}^2\rangle
  \ +\ 
 {1\over 8\gamma^2}
 \left[ 1\ +\ \gamma\rho_d\ +\ \gamma^2\rho_d^2\right]
   \ +\ {1\over 32 M_N^2}
   \nonumber\\
& = & 0.62\ \ \ \  +\ \ \ \ 2.33\ +\ 0.95\ +\ 0.39 \ +\   0.0014
\nonumber\\
& = & 4.30~{\rm fm}^2
\ \ \ \ ,
\label{eq:EFTcr}
\end{eqnarray}
where the last term is the relativistic correction.
Taking the square root of this value gives
$\sqrt{\langle r_d^2\rangle}=2.07~{\rm fm}$,
which is within a few percent of the measured value of
$\sqrt{\langle r_d^2\rangle}=2.1303~{\rm fm}$\cite{Wong,buch,FMS,EWa}.

%
\begin{figure}[h]
\qquad\qquad\qquad\psfig{figure=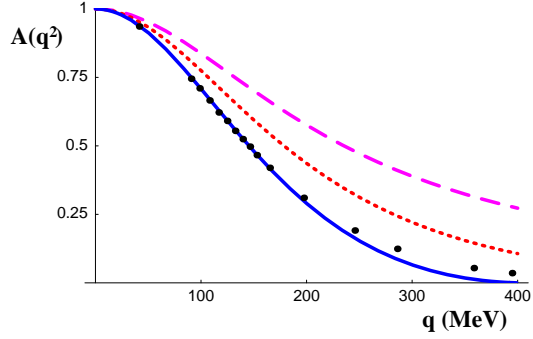,height=1.8in}
\noindent
\caption{\it The form factor $A( q^2)$ as  a function of 
$|{\bf q}|=\sqrt{-q^2}$.  
The dashed curve corresponds to the LO prediction,
the dotted curve corresponds to the NLO prediction,
and the solid curve corresponds to the NNLO 
prediction, in $\nopi$.
}
\label{fig:Aplot}
\vskip .2in
\end{figure}

At NLO the deuteron magnetic moment is 
\begin{eqnarray}
  \mu_d & = & {e\over 2 M_N} \left(\kappa_p\ +\ \kappa_n\ +\ \Ltwo\ { 2 M_N \gamma
    (\mu-\gamma)^2\over\pi}\right) 
  \ \ \ ,
\label{eq:deutmag}
\end{eqnarray}
the same expression as is found in the theory with pions.
Comparing the numerical value of this
expression with the measured value of $\mu_d$ gives (at NLO)
\begin{eqnarray}
  \Ltwo (m_\pi) & = & -0.149\ {\rm fm^4}
\ \ \ \ ,
\end{eqnarray}
at the renormalization scale $\mu=m_\pi$.
The evolution of the $\Ltwo (\mu)$ operator as the renormalization scale is
changed is determined by the RG equation
\begin{eqnarray}
    \mu {d\over d\mu}
    \left[ { \Ltwo \over \left(\Czerominus\right)^2} \right] 
     & = & 0
\ \ \ \ .
\label{eq:L2RG}
\end{eqnarray}

\begin{figure}[h]
\qquad\qquad\qquad\psfig{figure=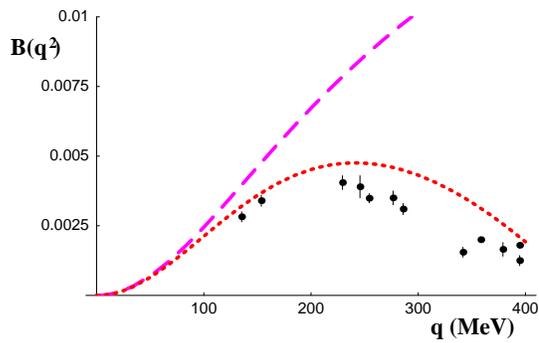,height=1.8in}
\noindent
\caption{\it The form factor $B( q^2)$ as  a function of 
$|{\b q}|=\sqrt{-q^2}$.  
The dashed curve corresponds to the LO prediction,
 and the dotted curve corresponds to the NLO prediction, 
 in $\nopi$.
}
\label{fig:Bplot}
\vskip .2in
\end{figure}

It is combinations of the electric, magnetic and quadrupole form factors that
are measured in elastic electron-deuteron scattering.
The differential cross section for unpolarized
elastic electron-deuteron scattering is
given  by
\begin{eqnarray}
{{\rm d}\sigma\over {\rm d}\Omega} = {{\rm
d}\sigma\over {\rm d}\Omega}\biggl\vert_{\rm Mott}\biggr.\[A(q^2) +
B(q^2)\tan^2\left({\theta\over 2}\right)
\]
 \ \ \ \ ,
\label{eq:eddiff}
\end{eqnarray}
where $A$ and $B$ are related to the form factors\cite{Wong} by
\begin{eqnarray}
A & = & F_C^2 + {2\over 3}\eta  F_M^2 + {8\over 9}\eta^2 F_{\cal Q}^2
\ \ \ \  ,
\nonumber\\
B & = & {4\over 3}\eta(1+\eta) F_M^2\ ,
\label{eq:ABdef}
\end{eqnarray}
with $\eta = -q^2/4M_d^2$.  
In order to compare with data, we take our
analytic results for the form factors and expand the expression 
eq.~(\ref{eq:ABdef}) in powers of $Q$.
At the order we are working, 
$A$ is sensitive both the electric and magnetic
form factors, 
while $B$ depends only on the magnetic form factor.
The predictions for $A(q^2)$ and $B(q^2)$ along with data are shown
in fig.~(\ref{fig:Aplot}) and fig.~(\ref{fig:Bplot}), respectively.

%

\begin{figure}[t]
\qquad\qquad\qquad\psfig{figure=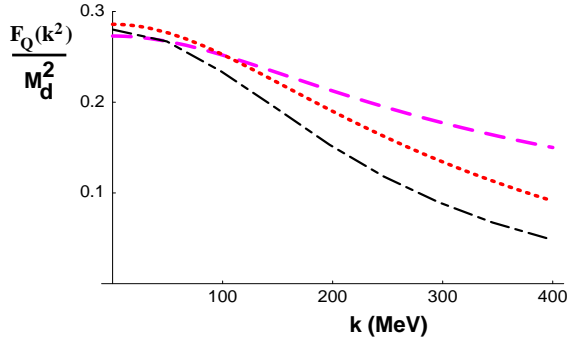,height=1.8in}
\noindent
\caption{\it The quadrupole form factor $F_{\cal Q}( |{\bf k}|^2)$ 
as  a function of  $|{\bf k}|$.  
The dashed curve corresponds to the LO prediction,
and the dotted curve corresponds to the NLO prediction, 
in $\nopi$.
The dot-dashed curve corresponds to a calculation with the Bonn-B potential
in the formulation of  \protect\cite{AGA} .
}
\label{fig:Qplot}
\vskip .2in
\end{figure}

Numerically, one finds that the quadrupole moment is given by
the sum of $ \mu_{\cal Q}^{(0)}  =  0.273\ {\rm fm^2}$ and
$ \mu_{\cal Q}^{(1)}  =  \left( 0.160\
  - \ 0.0165\ \CQuad \right) \ {\rm fm^2}$
giving, at NLO, 
\begin{eqnarray}
\mu_{\cal Q} & = & \left( 0.433\ -\ 0.0165\ \CQuad \right) \ {\rm fm^2}
\ \ \ ,
\end{eqnarray}
where  $\CQuad$ is the coefficient of a four-nucleon-one-photon operator, and 
is measured in ${\rm fm^5}$. Setting
$\CQuad=0$, we find a quadrupole moment of $0.433\ {\rm fm^2}$,
which is to be compared with the measured value of $0.287\ {\rm fm^2}$.
In order to reproduce the measured  value of the quadrupole moment
$\CQuad =+8.9\ {\rm fm^5}$, at $\mu=m_\pi$.
The quadrupole form factor resulting from this fit is shown in
fig.~(\ref{fig:Qplot}).

%


\subsubsection{$np\rightarrow d\gamma$}
 
In $\nopi$ we find a cross section for $np\rightarrow d\gamma$,
at NLO and at $|{\bf v}|=2200~{\rm m/s}$ incident neutron speed, of
\begin{eqnarray}
\sigma_{\pislashsmall} & = & \left( 287.1\ +\ 6.51\ \Lone\ \right)\ {\rm mb}
\ \ \ \ ,
\label{eq:npsig}
\end{eqnarray}
where  $\Lone$ is the coefficient of a four-nucleon-one-magnetic-photon operator,
with units of ${\rm fm}^4$ and is renormalized $\mu=m_\pi$. 
Requiring  $\sigma_{\pislashsmall}$ to reproduce the measured cross section 
$\sigma^{\rm expt}$ fixes $\Lone = 7.24~{\rm fm}^4$.

We see that even in the theory without dynamical pions, one is able to recover
the cross section for radiative neutron capture at higher orders.
It is clear that in this theory the four-nucleon-one-photon operators play a
central role in reproducing the low energy observables.
In the theory with pions, one can see by examining the contributing
Feynman diagrams\cite{SSWst},
that in the limit that the momentum transferred to the photon is small the
pion propagators can be replaced by $1/m_\pi^2$, while keeping the derivative
structure in the numerator.
This contribution, as well as the contribution from all hadronic exchanges,
is reproduced order by order in
the momentum expansion by the contributions from local multi-nucleon-photon
interactions.
From the calculations in the theory with dynamical pions, the value of $\Lone$
is not saturated by pion exchange currents as these contributions are
divergent, and require the presense of the $L_1$ operator\cite{SSWst}.
Therefore, estimates of $\Lone$ based on meson exchanges alone are
model dependent.

The effective range calculation of $np\rightarrow d\gamma$ was first performed
by Bethe and Longmire\cite{ERtheory} and revisited by Noyes\cite{Noyes}.
After correcting the typographical errors in the
expression for $\sigma$ that appears in
the Noyes article, the expressions in the two papers\cite{ERtheory,Noyes}
are identical,
\begin{eqnarray}
  \sigma^{( {\rm ER})} & = &
  {2\pi\alpha  \ \kappa_1^2 \ \gamma^6\ (a^{(\si)})^2\  a^{(\siii)}
    \over |{\bf v}|
    M_N^5 \left( 2\ -\ \gamma a^{(\siii)}\right)  }
  \nonumber\\
 & &  \left( 1\ +\ {1\over \gamma a^{(\siii)}}\ -\ {2\over\gamma a^{(\si)}} -
     {\gamma r_0^{(\si)}\over 2}  \right)^2
\ \ \ 
\label{eq:sigER}
\end{eqnarray}
which when expanded in powers of $Q$ is
\begin{eqnarray}
  \sigma^{( {\rm ER})} & = &
 {8\pi\alpha \kappa_1^2\gamma^3\over|{\bf v}|  M_N^5 } (1-\gamma a^{(\si)})
\left[
   (1-\gamma a^{(\si)})
\right.\nonumber\\
& & \left.
  \ +\
  {1\over 2}\gamma \left(\rho_d-r_0^{(\si)}\right) (1-\gamma a^{(\si)})
  \ +\
  {1\over 2}\gamma \left(\rho_d+r_0^{(\si)}\right)
  \ +\ ...\ 
\right]
\ \ \ .
\label{eq:sigERexp}
\end{eqnarray}

At LO in the $\nopi$ expansion the cross sections agree, however, at
NLO the expressions are very different.
In addition to the counterterm that appears at this order in the $\nopi$,
the contributions
from the effective range parameters are found to disagree, and more importantly
be renormalization scale dependent.
In the $\nopi$ the local counterterm is renormalized by the short-distance
behavior of graphs involving the $C_2$ operators and hence the effective range
parameters in both channels.  Given, this behavior it is no surprise that the
effective range contributions differ between the two calculations.
It is amusing to ask if there is a scale for
which
the expressions are identical, with $\Lone=0$.
Indeed such a scale exits,
\begin{eqnarray}
\mu^{{\rm ER}} & = &  {  \gamma  r_0^{(\si)}  a^{(\si)} - \rho_d\over
   a^{(\si)} \left( r_0^{(\si)} -  \rho_d\right)}
\ \ \  ,
\label{eq:musame}
\end{eqnarray}
which, by inserting the appropriate values, gives  scale $\mu^{{\rm ER}}\sim
144~{\rm MeV}$, coincidentally close to $\mu=m_\pi$.


\section{CONCLUSIONS}

We have performed a systematic analysis of the two-nucleon sector with
effective field theory using KSW power-counting and dimensional regularization.
It is clear that for each observable considered in this work
the perturbative expansion appears to be 
converging, and the expansion parameter(s)
is $\sim {1\over 3}$, both in the theory with pions and for the static
properties of the deuteron in the pionless theory.
I have deliberately avoided a discussion of the scale at which the theory with
dynamical pions breaks down, as this has been discussed in other
talks at this workshop\cite{NNsteele,NNkaplan,NNrupak,NNmehen}.
In the theory with pions, the NLO contributions from potential pion exchange
are found to be $\sim {1\over 3}$ the size of the NLO contributions from the
$C_2$ operators.
From these calculations, there is nothing to indicate that treating pions in
perturbation theory is not converging.
In order to be more confident that this is generally true,
higher order calculations in this theory must be performed.

As the pionless theory is very simple, computations beyond
NNLO will be performed, thereby giving calculations with better than $\sim 1\%$
precision.
As relativistic contributions to the deuteron static
properties are easy to calculate and are found to be
very small (suppressed by additional
factors of $m_\pi^2/M_N^2$) such high precision
calculations are possible in the near future.

\section{Acknowledgements}

I would like to thank the {\it Institute for Nuclear Theory }
for sponsoring
this workshop and particularly Wick Haxton who enabled us (the organizers) to
hold the type of workshop we envisaged.
I would also like to thank Maria Francom, Linda Vilet and Shizue Shikuma for
making the organization of the workshop pain-free.
Finally, I would like to thank my co-organizers, Paulo Bedaque, Ryoichi Seki and Bira
van Kolck, for a superb job.

\end{document}